\documentclass[aps,prl,reprint,superscriptaddress]{revtex4-1}

\usepackage{graphicx}% Include figure files
\usepackage{bm}% bold math
\usepackage{amsmath}
\usepackage{comment}

\begin{document}

%\preprint{}

\title{Quantum interference of resonance fluorescence from Germanium-vacancy color centers in diamond}


\author{Disheng Chen}
%\thanks{These two authors contributed equally}
\affiliation{Division of Physics and Applied Physics, School of Physical and Mathematical Sciences, Nanyang Technological University, Singapore 637371, Singapore}
\affiliation{The Photonics Institute and Centre for Disruptive Photonic Technologies, Nanyang Technological University, Singapore 637371, Singapore}

\author{Johannes E. Fr{\"o}ch}
\affiliation{School of Mathematical and Physical Sciences, University of Technology Sydney, Ultimo, NSW, 2007, Australia}

\author{Shihao Ru}
%\thanks{These two authors contributed equally}
\affiliation{Division of Physics and Applied Physics, School of Physical and Mathematical Sciences, Nanyang Technological University, Singapore 637371, Singapore}
\affiliation{Shaanxi Key Laboratory of Quantum Information and Quantum Optoelectronic Devices, School of Physics, Xi'an Jiaotong University, Xi'an 710049, China}

\author{Hongbing Cai}
%\thanks{These two authors contributed equally}
\affiliation{Division of Physics and Applied Physics, School of Physical and Mathematical Sciences, Nanyang Technological University, Singapore 637371, Singapore}
\affiliation{The Photonics Institute and Centre for Disruptive Photonic Technologies, Nanyang Technological University, Singapore 637371, Singapore}

\author{Naizhou Wang}
%\thanks{These two authors contributed equally}
\affiliation{Division of Physics and Applied Physics, School of Physical and Mathematical Sciences, Nanyang Technological University, Singapore 637371, Singapore}
\affiliation{The Photonics Institute and Centre for Disruptive Photonic Technologies, Nanyang Technological University, Singapore 637371, Singapore}

%\thanks{}

\author{Giorgio Adamo}
\affiliation{The Photonics Institute and Centre for Disruptive Photonic Technologies, Nanyang Technological University, Singapore 637371, Singapore}

\author{John Scott}
\affiliation{School of Mathematical and Physical Sciences, University of Technology Sydney, Ultimo, NSW, 2007, Australia}
\affiliation{ARC Centre of Excellence for Transformative Meta-Optical Systems (TMOS), Faculty of Science, University of Technology Sydney, Ultimo, New South Wales 2007, Australia}

\author{Fuli Li}
%\thanks{These two authors contributed equally}
\affiliation{Shaanxi Key Laboratory of Quantum Information and Quantum Optoelectronic Devices, School of Physics, Xi'an Jiaotong University, Xi'an 710049, China}

\author{Nikolay Zheludev}
\affiliation{Division of Physics and Applied Physics, School of Physical and Mathematical Sciences, Nanyang Technological University, Singapore 637371, Singapore}
\affiliation{The Photonics Institute and Centre for Disruptive Photonic Technologies, Nanyang Technological University, Singapore 637371, Singapore}
\affiliation{Optoelectronics Research Centre, University of Southampton, Hampshire, SO17 1BJ, UK}

\author{Igor Aharonovich}
\email[]{igor.aharonovich@uts.edu.au}
\affiliation{School of Mathematical and Physical Sciences, University of Technology Sydney, Ultimo, NSW, 2007, Australia}
\affiliation{ARC Centre of Excellence for Transformative Meta-Optical Systems (TMOS), Faculty of Science, University of Technology Sydney, Ultimo, New South Wales 2007, Australia}

\author{Wei-bo Gao}
\email[]{wbgao@ntu.edu.sg}
\affiliation{Division of Physics and Applied Physics, School of Physical and Mathematical Sciences, Nanyang Technological University, Singapore 637371, Singapore}
%\affiliation{MajuLab, CNRS-Université Côte d'Azur-NUS-NTU-SU International Joint Research Unit, UMI 3654, Singapore}
\affiliation{The Photonics Institute and Centre for Disruptive Photonic Technologies, Nanyang Technological University, Singapore 637371, Singapore}


\date{\today}

\begin{abstract}
Resonance fluorescence from a quantum emitter is an ideal source to extract indistinguishable photons. By using the cross polarization to suppress the laser scattering, we observed resonance fluorescence from GeV color centers in diamond at cryogenic temperature. The Fourier-transform-limited linewidth emission with $T_2/2T_1\sim0.86$ allows for two-photon interference based on single GeV color center. Under pulsed excitation, the 24 ns separated photons exhibit a Hong-Ou-Mandel visibility of $0.604\pm0.022$, while the continuous-wave excitation leads to a coalescence time window of 1.05 radiative lifetime. Together with single-shot readout of spin states, it paves the way towards building a quantum network with GeV color centers in diamond.
\end{abstract}


% \pacs{78.67.Hc,78.55.Cr,78.47.-p}

%\keywords{}

\maketitle


%\section{introduction}
Indistinguishable photons are indispensable resources for photonic quantum information processing \cite{flamini_photonic_2018} and underlie several key quantum technologies including linear optical quantum computing \cite{slussarenko_photonic_2019}, remote quantum-state teleportation \cite{pirandola_advances_2015}, and quantum-repeater-enabled large-scale quantum network \cite{kimble_quantum_2008}. Various optical processes or single-photon emitters have been explored to generate these identical photons, such as non-linear down-conversion process \cite{pan_multiphoton_2012}, single atoms \cite{reiserer_cavity-based_2015} or ions \cite{meraner_indistinguishable_2020}, semiconductor quantum dots \cite{schimpf_quantum_2021}, and solid-state quantum emitters \cite{awschalom_quantum_2018}. The latter stands out for the spin-tagged photonic interface \cite{hepp_electronic_2014}, mature nanostructure fabrications \cite{castelletto_advances_2017}, and the potential to scale up with the quantum photonic integrated circuits \cite{pelucchi_potential_2021, wan2020large}.

The negatively charged Germanium vacancy (GeV$^{-}$) color center in diamond exhibits a stable spectrum with negligible inhomogeneous broadening \cite{siyushev_optical_2017} thanks to the inversion-symmetry of its D$_{3d}$ molecular structure \cite{iwasaki_germanium-vacancy_2015}, which effectively suppresses the first-order response to the electric-field jittering \cite{rogers_multiple_2014}. Together with the high quantum efficiency of radiative decay (30\%) \cite{bhaskar_quantum_2017} and large zero-phonon line (ZPL) proportion (70\%) \cite{palyanov_germanium:_2015}, GeV color center presents a unique opportunity to realize solid-state quantum nodes without invoking any frequency-tuning technique \cite{sipahigil_indistinguishable_2014}.

%\section{here we show}
Here, we show that the presence of a microstructure around the GeV does not impair its optical properties and the lifetime-limited linewidth emission can be observed. This narrow linewidth allows for two-photon interference (TPI) based on indistinguishable photons from a single GeV emitter with a Hong-Ou-Mandel (HOM) effect \cite{hong_measurement_1987}. 
%of 60\%. Meanwhile, the enhanced photon collection efficiency by the solid-immersion lens (SIL) enables single-shot readout of GeV spin state, paving 
Together with the single-shot readout of GeV spin state enabled by enhanced photon collection efficiency using a solid immersion lens, this paves the road towards high-fidelity, high-rate spin-photon entanglement based on solid-state quantum emitters.

The GeV color centers in this work are generated via high-energy ion implantation (10 MeV, 10$^{10}$ cm$^2$) on a Type-IIa diamond substrate, followed by high-temperature high-vacuum annealing \cite{evans_narrow-linewidth_2016, chu_coherent_2014} that helps GeV formation and lattice repair (details in Supplemental Material \cite{GeV_supp_2022}). Before implantation, an array of solid immersion lens (SIL) is fabricated on the surface of the diamond via focused ion beam (FIB) milling, which provides a $3\sim8$ times boost in the photoluminescence (PL) collection efficiency as compared to the flat surface \cite{jamali_microscopic_2014, hadden_strongly_2010}. Acid treatment is applied to the sample before and after the annealing to ensure high-quality surface throughout the entire processing procedures \cite{sangtawesin_origins_2019}. In the end, each SIL contains multiple GeV color centers, and single quantum emitters can be selected via a combination of spatial mapping and resonant excitation thanks to a slight inhomogeneity of local strains around each GeV.

\begin{figure}[t]  
	\includegraphics{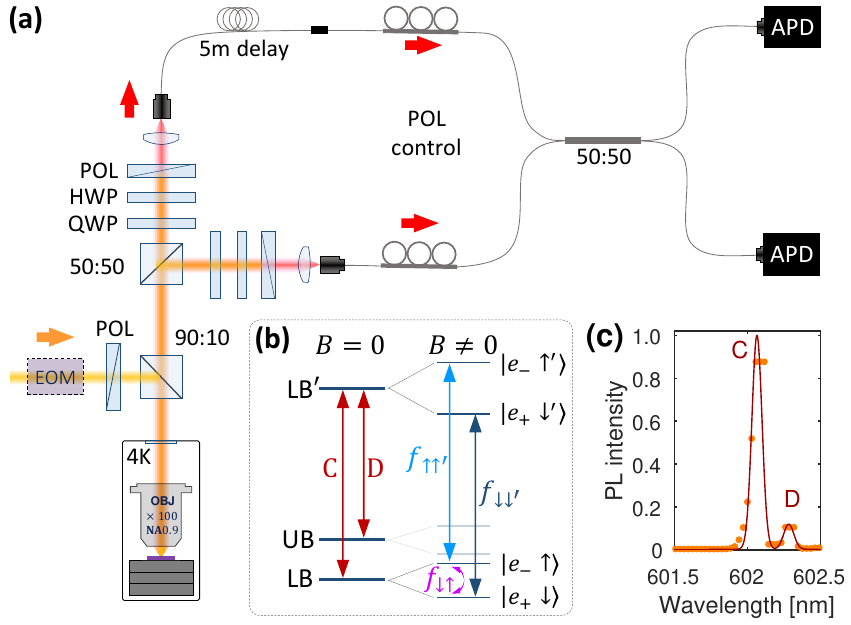}
	\caption{\label{fig:setup}  (a) Experiment setup. HWP: half wave-plate. QWP: quarter wave-plate. POL: polarizer. EOM: electro-optic modulator. APD: avalanche photodiode. (b) Energy levels of a GeV color center in diamond in zero and non-zero external magnetic field. The field lifts the double-degeneracy of the four orbitals, labeled as LB, UB, LB$^\prime$ and UB$^\prime$ (UB$^\prime$ is not shown) and reveals the spin degrees of freedom of the system including two cycling transitions $f_{\uparrow\uparrow^\prime}$ and $f_{\downarrow\downarrow^\prime}$ between LB and LB$^\prime$, and a flip-flop transition $f_{\uparrow\downarrow}$ in LB. LB: lower branch. UB: upper branch. Prime denotes the excited state. (c) Photoluminescence spectrum of GeVs at 4.2 K under 532 nm 0.46 mW excitation, monitored through a $600\pm7$ nm band-pass filter. The solid line is a Gaussian fit, finding the ground orbital splitting of 180 GHz between LB and UB.
	}
\end{figure}

The sample is cooled down to 4.2 K and interrogated using a home-built confocal microscope, as shown in Fig.~\ref{fig:setup}(a) (details in \cite{GeV_supp_2022}). When detecting resonance fluorescence from the GeV, the resonant laser scattering is suppressed via a cross polarization scheme at 30 dB extinction ratio. This is realized by adjusting the half wave-plate (HWP) and quarter wave-plate (QWP) in the collection path to tune the polarization of laser scattering perpendicular to the polarizer afterwards. When detecting phonon-side band (PSB) emission, a $650\pm20$ nm band-pass filter is placed in the collection path to reject the laser scattering with a suppression ratio of $>70$ dB.

\begin{figure}[t]
	\includegraphics{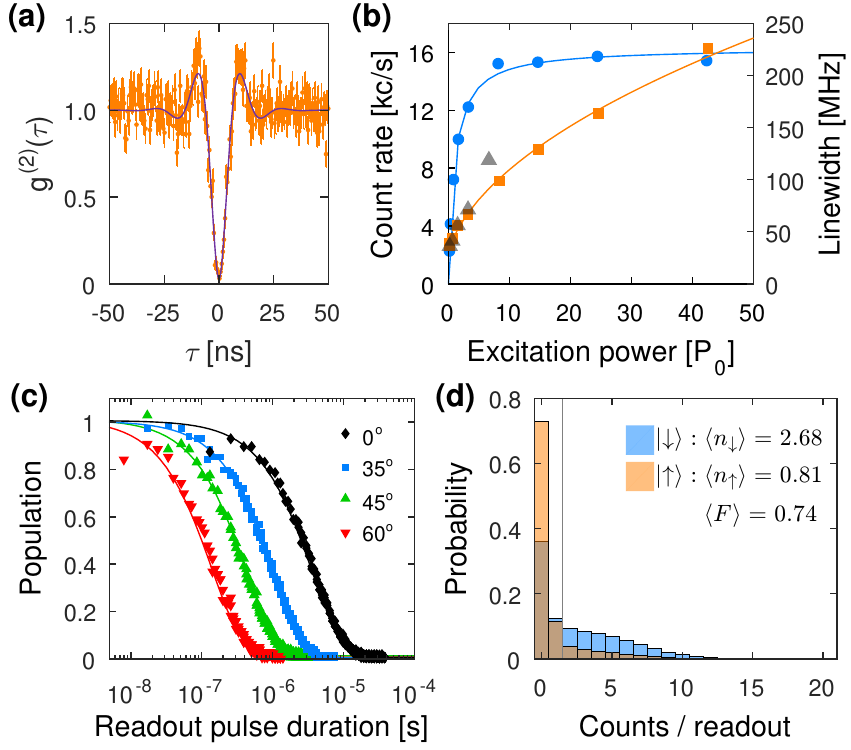}
	\caption{\label{fig:ple} (a) Second-order correlation function of PSB photons ($650\pm20$ nm) from a resonantly driven GeV. The solid line is a fit by solving a coherently driven 2-level system \cite{flagg2009resonantly}: $g^{(2)}(\tau) = 1 - \beta e^{-\eta\tau} [\cos(\nu\tau) + (\eta/\nu) \sin(\nu\tau)]$, with $\eta = (1/T_1 + 1/ T_2)/2$, $\nu = \sqrt{\Omega^2 - (1 / T_1 - 1 / T_2)^2/4}$. Here, $\beta = 0.972$ is the dip depth, $T_1=5.5$ ns and $T_2=7.1$ ns are the lifetime and coherence time of the excited state, and $\Omega = 0.57$ GHz is the Rabi frequency. 
	(b) Count rate (blue circle) and linewidth (orange square for PSB, grey triangle for ZPL) extracted from a Lorentzian fit to the PLE spectra in Supplemental Material. 
	%The orange rectangles and black triangles are the PLE linewidths for collecting PSB or ZPL photons, respectively. The blue circles are the resonant PSB intensity. 
	Solid curves are the 2-level model fitting, with the saturation power P$_0$ = 6.1 $\pm$ 0.7 nW and coherence time $T_2$ = 9.5 $\pm$ 0.4 ns. 
	(c) $\left|\downarrow \right\rangle$ population during the readout pulse that addresses $f_{\downarrow\downarrow^\prime}$ transition resonantly. Prior to the readout, the spin is initialized to $\left|\downarrow\right\rangle$ state by pumping $f_{\uparrow\uparrow^\prime}$ transition at 1.6 P$_0$. The magnetic field is held at 1.107 T for all measurements; only the direction varies. Solid lines are single exponential fits with time constants 3.8 $\mu$s, 1.0 $\mu$s, 0.4 $\mu$s, and 0.15 $\mu$s for 0$^\circ$, 35$^\circ$, 45$^\circ$, and 60$^\circ$ orientated field with respect to the sample plane, respectively. (d) Photon statistics of single-shot readout when reading $\left|\downarrow\right\rangle$ (blue) or $\left|\uparrow\right\rangle$ (red) state in $\rm B=1.107$ T along 0$\deg$. $\langle n_{\uparrow} \rangle$ and $\langle n_{\downarrow} \rangle$ are the average readout photon numbers. $\langle \mathcal{F} \rangle$ is the average fidelity.  
	}
\end{figure}

When the GeV is illuminated with 532 nm non-resonant light, even though both excited orbitals are equally populated, the population in the upper branch (UB$^\prime$) immediately relaxes to the lower branch (LB$^\prime$) by emitting phonons [see Fig.~\ref{fig:setup}(b)]. The lack of high-energy phonons ($\sim$1 THz) in diamond at 4 K allows for the accumulation of almost all population in LB$^\prime$ before any radiative decay takes place \cite{jahnke_electronphonon_2015}. This results in a two-line structure in the PL spectrum \cite{ekimov_germaniumvacancy_2015}, corresponding to the decays from LB$^\prime$ to the double-degenerated ground orbitals LB and UB, as shown in Fig.~\ref{fig:setup}(c). In fact, multiple GeVs are present in the excitation volume and contribute to the observed two-line structure. Thanks to the narrow emission linewidth and slight variations of local strain environment around each emitter, single GeVs can be addressed by employing resonant excitation, as shown by the wide-range photoluminescence excitation (PLE) spectrum in \cite{GeV_supp_2022}. We focus on the GeVs with bright and stable emission for further studies. By monitoring the PSB emission from the resonantly addressed GeV, we are able to confirm the singleness of the photon source by measuring the second-order correlation function in Hanbury Brown-Twiss (HBT) configuration \cite{hanbury_brown_test_1956}, typically observing a value $g^{(2)}(0)=0.028\pm0.009$, as shown in Fig.~\ref{fig:ple}(a).

% PLE data description
To evaluate the dephasing of these optical transitions, we conduct power-dependent PLE measurements on the GeV by collecting either PSB  or ZPL emission. The increase of resonant excitation power broadens the PLE linewidth evidently and saturates the emission intensity at P$_0=6.1$ nW, matching the predictions of a 2-level system, as shown in Fig.~\ref{fig:ple}(b). Apart from the deviation of ZPL PLE linewidth at 6 P$_0$, caused by the fluctuations of resonant laser scattering, the main difference is the 3 times stronger intensity for resonance fluorescence than the PSB emissions thanks to the 2:1 ZPL/PSB ratio and the finite PSB detection bandwidth (defined by the filter $\sim40$ nm). Extrapolating the excitation power to 0 nW, we obtain an optical linewidth of 34 MHz, corresponding to a coherence time of T$_2=9.5$ ns. Considering the excited state lifetime T$_1$ of 5.5 ns, determined by time-resolved measurements \cite{GeV_supp_2022}, we obtain $T_2/2T_1=0.86$, marking the Fourier-transform limited linewidth emission from a GeV. We confirm this narrow linewidth emission on several GeVs located in different SILs \cite{GeV_supp_2022}. The generality of this excellent optical properties across the sample implies that our treatment of the diamond, including high-temperature annealing and acid cleaning, are beneficial to stabilize the local environment around GeV color centers.

According to the PL spectrum in Fig.~\ref{fig:setup}(c), this GeV possesses a ground-state splitting of 180 GHz, which is 20 GHz greater than the intrinsic non-strained value of 160 GHz \cite{bhaskar_quantum_2017}, and can be categorized as a low-strain environment. Since moderate strains cannot dominate the spin-orbit coupling and the different coupling strengths give rise to different $g$-factors for the ground and excited states \cite{nguyen_integrated_2019}, an external magnetic field thus is able to produce two spin-selective transitions, $f_{\downarrow\downarrow^\prime}$ and $f_{\uparrow\uparrow^\prime}$, as shown in Fig.~\ref{fig:setup}(b). These spin conserving transitions are ideal for single-shot readout of spin states. Due to the different anisotropies of $g$ tensors of the ground and excited states, the number of repetitive readout until a spin flip is field-orientation dependent, as shown in Fig.~\ref{fig:ple}(c). Generally, a larger field misalignment from the symmetry axis of GeV induces a faster spin relaxation. Experimentally, we initialize the system to $\left|\downarrow\right\rangle$ state by pumping $f_{\uparrow\uparrow^\prime}$ transition for 500 $\mu$s followed by a 1 ms readout that addresses $f_{\downarrow\downarrow^\prime}$ transition. When the magnetic field is closely aligned to the GeV symmetric axis, the spin can withstand a thousand times readout before experiencing a flip. By selecting an optimum readout window of 80 $\mu$s and a threshold of 1.5 photons per readout \cite{GeV_supp_2022}, we achieve single-shot readout of $\left| \uparrow \right\rangle$ and $\left|\downarrow\right\rangle$ state with a fidelity of 63.9\% and 84.5\%, respectively, as shown in Fig.~\ref{fig:ple}(d). The final fidelity $\mathcal{F}$ of readout, i.e., the average of the two, is 74.2\%, limited by the spin pumping efficiency of initialization \cite{GeV_supp_2022}.

\begin{figure}[t]
	\includegraphics{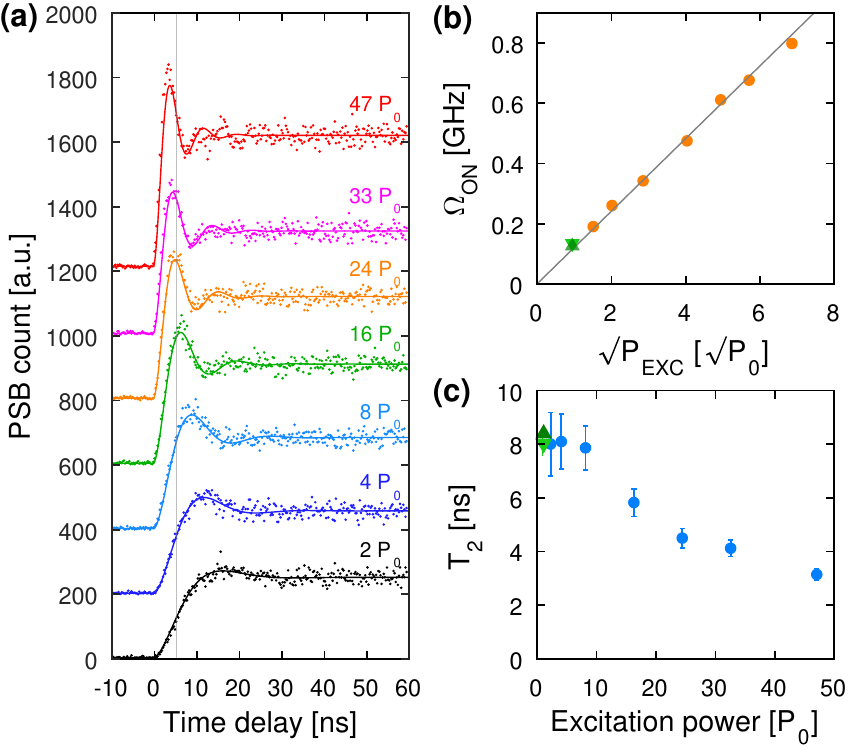}
	\caption{\label{fig:rabi}  (a) Optical Rabi oscillations of the GeV under resonant excitations (vertical shifted for clarity). The solid lines are the fits by solving semi-classical 2-level master equation \cite{GeV_supp_2022}. The vertical grey line marks the 5-ns pulse width used for pulsed two-photon interference (TPI) experiment. (b) Rabi frequency $\Omega_\text{ON}$ when the EOM switches on, following linearly over the square root of excitation power (grey). (c) Coherence time $T_2$ extracted from the Rabi oscillations in (a). The green upward and downward triangles in (b) and (c) are the parameters extracted from the continuous-wave TPI measurements. }
\end{figure}

To find the optical $\pi$-pulse of resonant excitation, we study the power-dependent time-resolved PL from the GeV by modulating the excitation beam with an electro-optic modulator (EOM), as shown in Fig.~\ref{fig:rabi}(a).  We model the optically driven GeV as a two level system using the master equation with Lindblad terms that consider both spontaneous decay and pure dephasing \cite{GeV_supp_2022}. The extracted Rabi frequency $\Omega_\text{ON}$ (when the EOM switches on) increases linearly over the square root of the excitation power [Fig.~\ref{fig:rabi}(b)],while the coherence time $T_2$ drops monotonically from 8 ns to 3 ns as the excitation power increases from 2 P$_0$ to 50 P$_0$ [Fig.~\ref{fig:rabi}(c)]. We tentatively attribute the escalated dephasing to the laser-induced environmental fluctuations \cite{chen_optical_2019}. Two factors are considered for optimizing the pulse duration: a short pulse is needed to minimize the two-photon emission probability during the excitation period, while a long pulse is favored for laser suppression. We choose a pulse length close to the excited state lifetime with a power of 39 P$_0$ to realize the $\pi$-rotation of orbital populations. 

\begin{figure}[t]
	\includegraphics{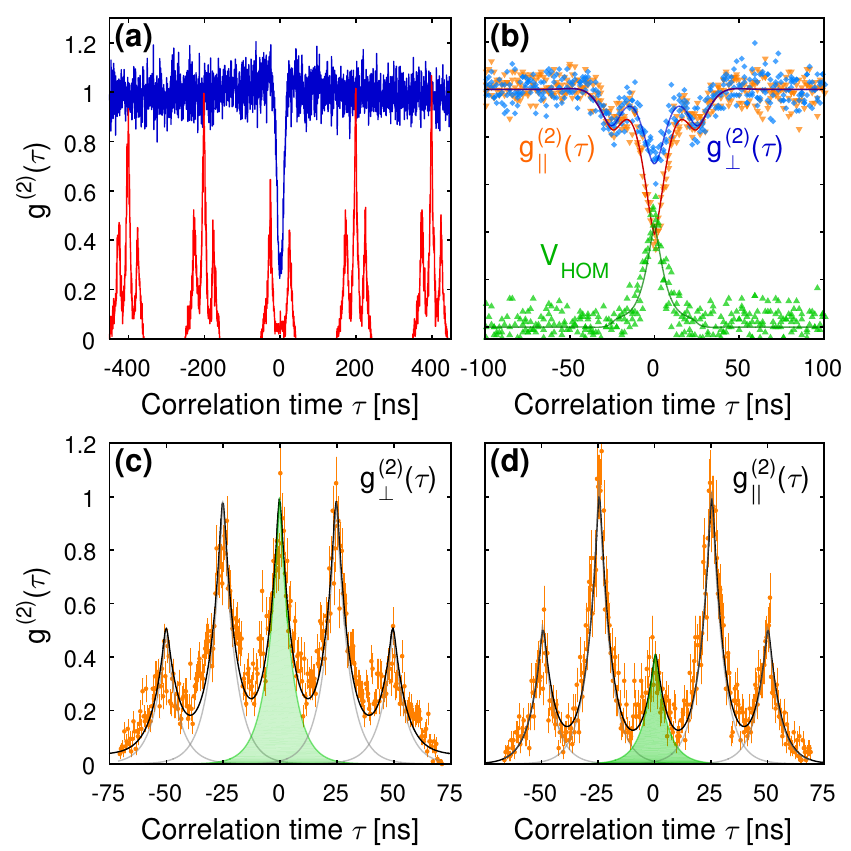}
	\caption{\label{fig:tpi}  (a) Normalized HBT correlations of ZPL photons from the CW-driven (blue) or pulsed-driven (red) GeV, with resonant power 0.8 P$_0$ and 39 P$_0$, respectively. The three correlation peaks in pulsed excitation correspond to the two consecutive excitations separated by $\delta t$. (b) CW TPI when detecting ZPL photons in aligned ($g_\parallel^{(2)}(\tau)$, orange) or orthogonal ($g_\perp^{(2)}(\tau)$, blue) polarizations. The solid lines are the fitting by solving 2-level master equation. V$_{HOM}$ is the Hong-Ou-Mandel visibility following $1-g_\parallel^{(2)}(\tau)/g_\perp^{(2)}(\tau)$. (c-d) Pulsed TPI results measured in cross (c) or parallel (d) polarizations. The black curves are the sum of five exponential cusps (grey). The areas of the central shaded green cusp are used to evaluate the HOM visibility. See the main text for details. }
\end{figure}

Due to electron-phonon interactions \cite{jahnke_electronphonon_2015}, the PSB photons are distributed across a wide spectral range and are distinguishable in energy. Therefore, the coherent ZPL photons have to be used for TPI measurements. But the resonant laser scattering is not completely suppressed due to the finite suppression ratio, which can introduce a non-trivial impact on the photon statistics. We evaluate this influence by measuring HBT statistics of ZPL emissions from the GeV, as shown in Fig.~\ref{fig:tpi}(a). Although the pulsed excitation power is 50 times stronger than that of the CW driving, the deterioration of the anti-bunching dip at $\tau = 0$ is more prominent for the latter than the former (0.3 vs 0.05). This may relate to the dark state of the GeV color center \cite{GeV_supp_2022}. 
%which effectively reduces the signal-background ratio, leading to a rising of $g^{(2)}(0)$. Earlier studies show that the CW resonant driving can shelve the GeV emitter into a dark state via its excited state over a time scale of ms \cite{chen_optical_2019}. In pulsed excitation, however, each excitation sustains only a short period of time of 5 ns ($\sim T_1$) while the dwell time between two consecutive excitations is more than $5~T_1$. This waiting period is long enough for the GeV to relax back to its ground state via radiative decay, and prevents itself from being shelved to the dark state by the second excitation.

To perform TPI, we delay one emitted photon, and interfere two consecutive photons emitted from the same defect at the beam splitter. The delay $\delta t$ is about 25 ns, achieved by adding an additional 5-meter-long optical fiber to one detection arm, as shown in Fig.~\ref{fig:setup}(a). This delay is almost twice the coherence time of single photons from the GeV color center and ensures the vanishing probability of self field-field interference at the BS. By controlling the polarizations of the interfering photons, we conduct TPI measurements for both indistinguishable [$g_\parallel^{(2)}(\tau)$] and distinguishable [$g_\perp^{(2)}(\tau)$] photons, as shown in Fig.~\ref{fig:tpi}(b). The deeper central dip of $g_\parallel^{(2)}(\tau)$ as compared to $g_\perp^{(2)}(\tau)$ reflects the TPI of indistinguishable photons, imposed by the bosonic nature of photons \cite{fearn_theory_1989}. The non-vanishing $g_\parallel^{(2)}(0)$, on the other hand, implies the imperfect experimental conditions including excitation laser leakage and dephasing of the photon source over time $\delta t$. We note that the instrument response function (IRF) here is at least one order of magnitude faster than the dip width of the correlation functions, thus playing a negligible role in data processing.

The HOM visibility $V_\text{HOM}(\tau)$ is evaluated via $V_\text{HOM}(\tau) = (g_\perp^{(2)}(\tau) - g_\parallel^{(2)}(\tau)) / g_\perp^{(2)}(\tau)$, as shown by the green triangles in Fig.~\ref{fig:tpi}(b). However, the value $V_\text{HOM}(0)$ strongly depends on the jittering of the detectors and a perfect detector with zero response time will always measure $V_\text{HOM}(0) = 0$ no matter how different the frequencies of two interfering photons are \cite{koong_fundamental_2019}. Thus, another figure-of-merit, the coalescence time window CTW$=\int d\tau V_\text{HOM}(\tau)$, \cite{proux_measuring_2015} is employed to quantify the indistinguishability. This value defines a time window beyond which no more deterministic coalescence of two photons can take place at the BS. Our data gives a CTW of 5.8 ns, close to the excited state lifetime $T_1$ of the GeV. Ideally, it should be about 2 $T_1$ for perfectly indistinguishable photons, whereas in our case, the dephasing ($T_2 < 2 T_1$) and the residual laser photons compromise this figure. We model the GeV system as a coherently driven 2-level emitter plus a resonant laser background to fit the measured results \cite{GeV_supp_2022}, and find a Rabi frequency $\Omega=0.13$ GHz and a coherence time $T_2$ of $8.4\pm0.2$ ns and $8.0\pm0.4$ ns for $g_\parallel^{(2)}(\tau)$ and $g_\perp^{(2)}(\tau)$, respectively. All these numbers are consistent with the earlier Rabi measurements, as shown by the green triangles in Fig.~\ref{fig:rabi}(b) and (c).

We also evaluate the indistinguishability of pulse-excitation generated single photons since these photons can be produced on demand. By matching the separation of two excitation pulses to the path difference of two detection arms, we observe five correlation peaks at the correlation time $\tau = -2 \delta t$, $-\delta t$, 0, $\delta t$,  and $2 \delta t$, respectively, with an amplitude ratio of 1:2:2:2:1 if the photons are distinguishable, as shown in Fig.~\ref{fig:tpi}(c). After aligning the polarizations, the TPI reduces the coincidence count at $\tau$=0, leading to an amplitude ratio of 1:2:x:2:1, as shown in Fig.~\ref{fig:tpi}(d). The incomplete vanishing of the central peak implies the distinguishable properties of interfering photons inherited from the photon source and the potential contamination from the residual laser photons. Phenomenologically, we fit each peak as a cusp of single exponential decay following $A \exp (-\left|\tau-t_0 \right| / \tau_0)$, where $\tau_0$ describes the time span of interfering photons and is shared among all ten peaks, and amplitude A represents the area of each peak that follows the ratio above. The HOM visibility can be evaluated via $V_\text{HOM} = (A_\perp-A_\parallel)/A_\perp=0.604\pm0.022$, with $A_\perp$ and $A_\parallel$ representing the areas of the central cusp (green shaded region) in Fig.~\ref{fig:tpi}(c) and (d), respectively. This visibility is consistent with the fitting result based on a semiclassical model \cite{GeV_supp_2022}.

To improve the TPI visibility, we have to increase the laser suppression ratio and alleviate the dephasing of the photon source. Regarding the suppression, a few orders of magnitude improvement is possible \cite{kuhlmann_dark-field_2013} if the sample vibrations can be restrained \cite{GeV_supp_2022}. Alternatively, switching to other nanostructures for collection enhancement, such as nanobeam photonic structure \cite{burek_high_2014} or nano pillars \cite{zhang_complete_2017} may also alleviate the issue. Despite the symmetry-protected optical transitions, we still observe spectral diffusions of GeVs over days, possibly caused by the second-order Stark effect and local strain fluctuations \cite{meesala_strain_2018}.  One solution is to exploit Purcell effect of nanocavities to broaden the emission line to exceed the spectral diffusion \cite{evans_photon-mediated_2018}. Alternatively, one can utilize strain tuning techniques to actively counter the spectral diffusion, provided a high-enough collection efficiency to enable transition frequency determination on a rate faster than the spectral diffusion \cite{machielse_quantum_2019, sohn_controlling_2018}.

% Conclusions:
In conclusion, we demonstrate lifetime-limited linewidth emission from the GeV color center in diamond with $T_2/2T_1\sim0.86$. The enhanced collection efficiency of SIL microstructure allows for single-shot readout of spin states of the GeV color center with a fidelity of 74\%, limited by the spin pumping efficiency of 80\%. This can be improved by carefully aligning the magnetic field to the symmetric axis of the GeV. The two 25 ns-separated ZPL photons from a single GeV possess a coalescent time window of $\approx T_1$ under CW driving and a HOM visibility of $V_\text{HOM}=0.604\pm0.022$ under pulsed excitation. The TPI performance is currently limited by the finite laser suppression of resonant laser and local strain fluctuations. Utilizing the strain tuning technique to feedback stabilize the optical transition frequency, it is possible to overcome these limitations. 
%This work paves the road towards realizing high-rate spin-photon entanglement based on GeV spin in diamond and lays down the foundation for entangling distant GeV-based solid state spins via photonic interface.        

\begin{acknowledgments}
We acknowledge Singapore National Research foundation through QEP grant (NRF2021-QEP2-01-P02, NRF2021-QEP2-03-P01, 2019-0643 (QEP-P2) and 2019-1321 (QEP-P3)) and Singapore Ministry of Education (MOE2016-T3-1-006 (S)), the Australian Research council (via CE200100010), the Asian Office of Aerospace Research and Development grant FA2386-17-1-4064.

\end{acknowledgments}

\nocite{*}

\end{document}

\title{Supplemental Material for\vspace{0.08cm}\\ 
		\small ``Quantum interference of resonance fluorescence from Germanium-vacancy color centers in diamond''}
	
	\author{{Disheng Chen$^{1, 2}$, Johannes Fr{\"o}ch$^3$, Shihao Ru$^{1, 4}$, Hongbing Cai$^{1, 2}$, Naizhou Wang$^{1, 2}$, Giorgio Adamo$^{2}$, John Scott$^{3}$, Fuli Li$^{4}$, Nikolay Zheludev$^{1,2,5}$, Igor Aharonovich$^{3, 6}$, Wei-bo Gao$^{1, 2}$\vspace{0.2cm}\\}
		{\small $^1$Division of Physics and Applied Physics, School of Physical and Mathematical Sciences, Nanyang Technological University, Singapore 637371, Singapore\\	
			$^2$The Photonics Institute and Centre for Disruptive Photonic Technologies, Nanyang Technological University, Singapore 637371, Singapore\\
			$^3$School of Mathematical and Physical Sciences, University of Technology Sydney, Ultimo, NSW, 2007, Australia\\
			$^4$Shaanxi Key Laboratory of Quantum Information and Quantum Optoelectronic Devices, School of Physics, Xi'an Jiaotong University, Xi'an 710049, China\\
			$^5$Optoelectronics Research Centre, University of Southampton, UK\\
		    $^6$ARC Centre of Excellence for Transformative Meta-Optical Systems (TMOS), Faculty of Science, University of Technology Sydney, Ultimo, New South Wales 2007, Australia}}
	
	%\date{\today} % PRL won't change the date to match the eventual paper, so leave it out.
	%\pacs{78.67.Hc,78.55.Cr,78.47.-p}
	
	\maketitle
	
	\renewcommand\theequation{S\arabic{equation}}
	\renewcommand{\thefigure}{S\arabic{figure}}
	
	In this Supplementary Material, we provide further information on:\vspace{0.1cm}\\
	\hspace*{0.6cm}Section 1:  Sample fabrication  and setup details  \\
	\hspace*{0.6cm}Section 2:  Isolate single GeV color centers and laser suppression \\
	\hspace*{0.6cm}Section 3:  PLE, RPLE, lifetime $T_1$, and PLE linewidth of other GeVs\\
	\hspace*{0.6cm}Section 4:  Field-direction dependent PLE \\
	\hspace*{0.6cm}Section 5:  Single-shot readout – threshold $n_\text{th}$ and readout fidelity $\mathcal{F}$ \\
	\hspace*{0.6cm}Section 6:  Rabi oscillations – model \\
	\hspace*{0.6cm}Section 7:  Polarization alignment for TPI experiments \\
	\hspace*{0.6cm}Section 8:  TPI - continuous-wave excitation \\
	\hspace*{0.6cm}Section 9:  TPI - pulsed excitation \vspace{0.1cm}
	
	All figures and equations in Supplementary Material are labeled with the prefix ``S'' to distinguish from those in the body of the Letter.
	
	\begin{figure*}[b]
		\includegraphics{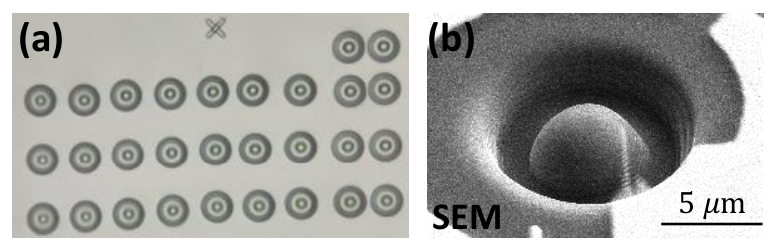}
		\caption{\label{fig:sil} (a) Optical image of an array of solid immersion lens (SIL) fabricated before ion implantations. All SILs contain GeV color centers after the implantation and annealing. (b) Scanning electron microscope (SEM) image of one SIL, showing a dimension of $5\mu$m wide $2.5 \mu$m deep. These nanostructures provide $>3$ times enhancement in collection efficiency compared to the flat surface.}
	\end{figure*}

	\section{SECTION 1: Sample fabrication \& setup details}
	
	Prior to fabrications, the Type-IIa electronic grade diamond substrate (Element Six) was cleaned in hot Piranha acid (150 $^\circ$C 3:1 H$_2$SO$_2$:H$_2$O$_2$) for 1 hr. After cleaning the substrate was sputter coated with a Gold/Palladium film of 10 nm thickness for charge mitigation during processing.  The microstructures were fabricated using FEI DB235 with a Ga+ ion beam of 30 kV primary energy and 7 nA beam current. The fabricated solid immersion lens has an approximate diameter of 5 $\mu$m and a height of 2.5 $\mu$m as shown in Fig.~\ref{fig:sil}(a) and (b). After fabrications, the Au/ Pd film was removed using hot Aqua Regia (100 $^\circ$C 3:1 HCl:HNO$_3$). Ge implantation was done with a charge state of 4, energy of 10 MeV, fluence of $10\times10^{10}$ at a current of 1.5 nA. After implantation, the diamond was again cleaned in hot Piranha Acid before placed in a furnace and annealed with the following protocol:\vspace{0.2cm}\\
	\hspace*{0.6cm}(1) Ramp to 100 $^\circ$C over 2 h. Hold for 11 h. \\
	\hspace*{0.6cm}(2) Ramp to 400 $^\circ$C over 4 h. Hold for 8 h. \\
	\hspace*{0.6cm}(3) Ramp to 800 $^\circ$C over 12 h. Hold for 8 h. \\
	\hspace*{0.6cm}(4) Ramp to 1100 $^\circ$C over 12 h. Hold for 8 h. \\
	\hspace*{0.6cm}(5) Cool down to room temperature. \vspace{0.2cm}
	
	The slow, long-term annealing helps GeV creations and stabilize the required -1 charge state by curing the lattice damages caused by the high-energy implantations. After annealing we removed a-few-nanometer carbonized surface layer by cleaning the sample in hot acid for 8 hours (210 $^\circ$C 1:1 H$_2$SO$_4$:HNO$_3$).   
	 
	The treated sample is mounted a XYZ piezo scanner that is attached to a XYZ piezo steppers for fine and coarse positioning GeV emitters with respect to the objective. The positioners and the objective are fixed on a home-built cage system that is attached to the bottom of MXC plate in a $^3$He/$^4$He dilution refrigerator (Bluefors LD250). The AMI 9-3T two-dimensional vector superconducting magnet allows arbitrary orientation of the magnetic field in YZ plane.  
	
	In our optical setup, the resonant excitation beam is reflected from a 90:10 beam splitter (BS) before reaching the objective (NA 0.9, WD 210 $\mu$m) to address the GeV of interest, as shown in Fig. 1(a) of the main text. The photoluminescence (PL) of GeVs is collected by a single-mode fiber and directed to a spectrometer for spectral characterization or to an avalanche photodiode (APD) for photon counting. Two detection arms are used for two-photon interference (TPI) measurements, where an extra 5 m-long optical fiber is inserted in one detection arm to provide the needed time delay. The half-wave plate (HWP) and quarter-wave plate (QWP) in the collection paths are set to establish cross polarization scheme to eliminate the resonant laser scattering at an extinction ratio of 30 dB, while the fiber-based polarization control in each arm is used to generate cross or parallel polarizations between the two interfering photons at the second BS. 
	%Throughout the experiments, the detection polarization on the cage system has been set to eliminate the reflections of resonant laser beam with an extinction ratio of 30 dB. 
	When phonon-side band (PSB) detection is needed, a $650\pm20$ nm band-pass filter is placed in collection path (between the two BS cubes) to reject the laser reflection at 70 dB suppression. 
	
	We note that a low-power green laser is always introduced to the GeV during resonant excitation. It is needed to stabilize the GeV's charge state, without which no resonance fluorescence of GeV can be detected. We stress that the laser power (of the green laser) is so weak that no fluorescence can be induced from the GeV by itself. The only function it serves is to tune the local Fermi level in favor of -1 charge state, as explained in earlier studies \cite{chen_optical_2019}.

	\begin{figure*}[b]
		\includegraphics[scale=0.9]{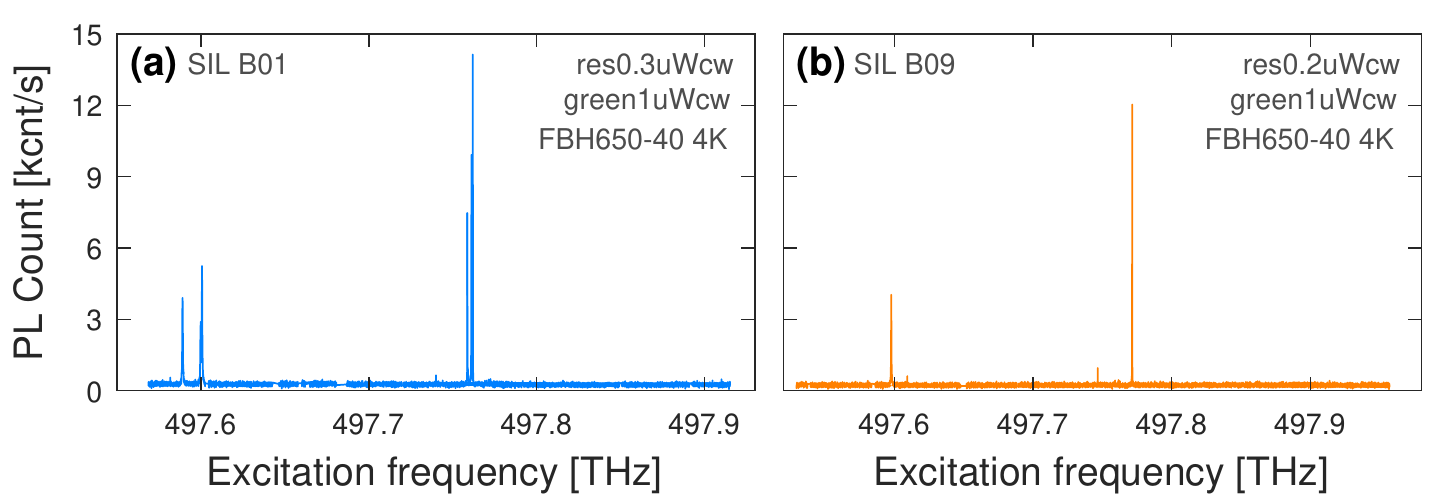}
		\caption{\label{fig:wide} (a) Wide-range PLE scan of two bright spots found in SIL B01 (a) and B09 (b), respectively. These bright spots are firstly found via PL spatial mapping under 532 nm non-resonant excitation at 0.5 mW. These two PLE spectra are taken with a $650\pm20$ nm band pass filter (PSB detection). Each discrete peak in the spectra corresponds a single GeV within the excitation volume.
		}
	\end{figure*}

	\section{SECTION 2: Isolate single GeV color centers \& laser suppression}
	
	To locate GeVs for resonant excitation, we firstly find their rough positions in a SIL by conducting confocal mapping under 532 nm non-resonant excitation (0.5 mW). Prior to the mapping, we set the laser focus to the depth of SIL. We monitor the zero-photon line (ZPL) emissions from GeVs by using a $600\pm7$ nm band-pass filter. This narrow bandwidth helps reduce the ``fake'' signals from other emitters or defects. After the scan, we identify each bright spot by its spectrum. 
	
	Next, we pinpoint the resonant frequencies of GeVs of a bright spot by conducting a 300-GHz-wide frequency scan, as shown in Fig.~\ref{fig:wide}(a) and (b). Each peak in the photoluminescence excitation (PLE) spectra corresponds to a single GeV emitter. The discreteness of the emission energy originates from the variations of local strain environment around each GeV. This allows us to resonantly address each GeVs and check the resonant excitation stability. We pick bright and stable emitters for further studies.   
	
\begin{figure*}[b]
		\includegraphics[scale=0.8]{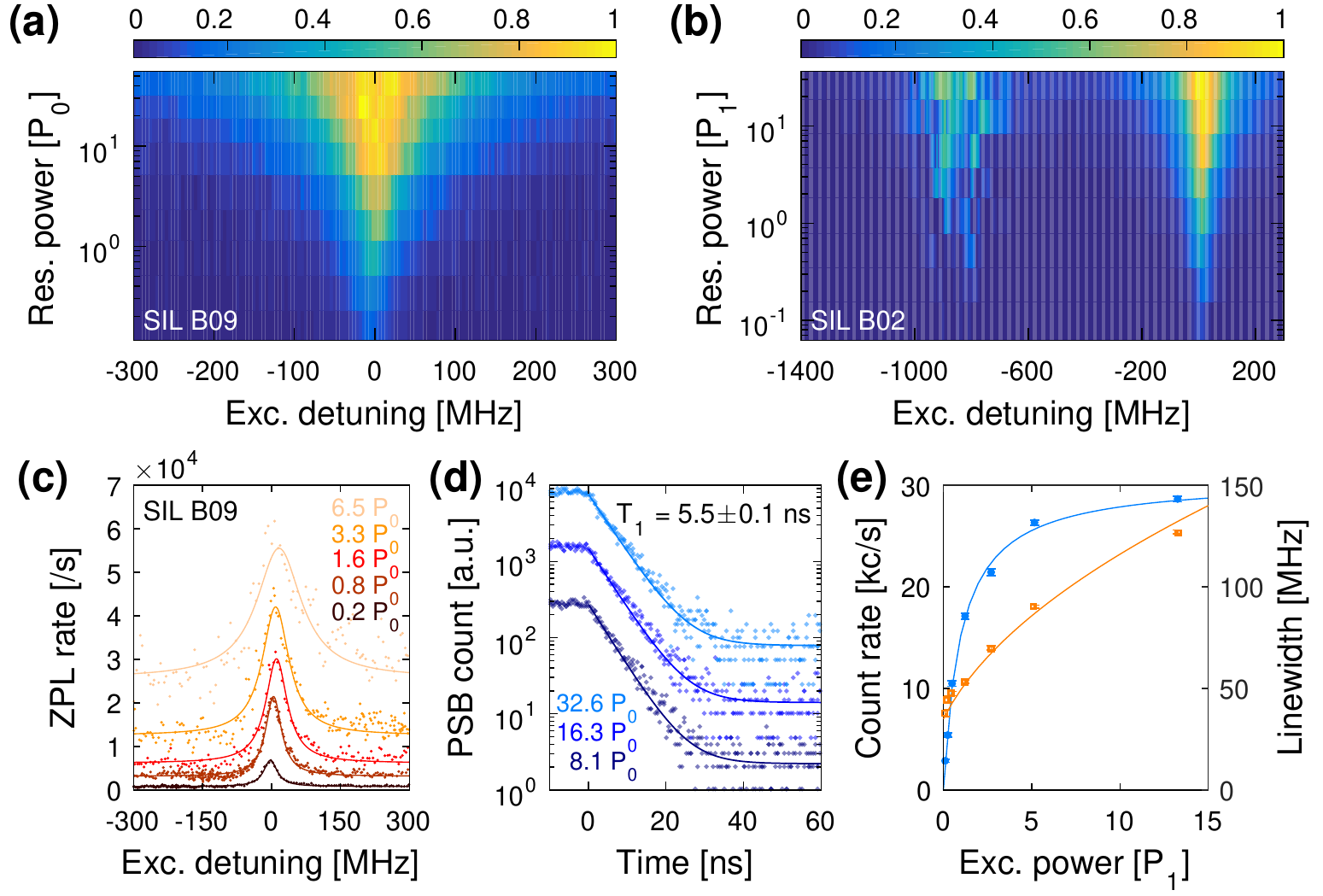}
		\caption{\label{fig:rple} (a) Power-dependent PLE spectra of the GeV color center studied in the main text (located in SIL B09), monitored through PSB emissions. The resonant power is labeled in the unit of saturation power P$_0$. (b) PLE spectra of another GeV in SIL B02. Multiple peaks appear in the spectra possibly stemming from the GeVs nearby. (c) RPLE spectra of the GeV in SIL B09 by monitoring its ZPL emission with a $600\pm7$ nm band-pass filter. Solid lines are the best fit of Lorentzian line. (d) Excited-state lifetime measured at three excitation powers (vertically shifted for clarity). The IRF is $\sim0.3$ ns, one order of magnitude shorter than $T_1$. Solid curves are the single exponential fit, with an average $T_1=5.5\pm0.1$ ns. (e) Extracted count rate (blue) and linewidth (orange) from the Lorentzian fitting in (b). The solid lines are the fitting based on a coherently driven 2-level system.
		}
	\end{figure*}

	To detect the resonance fluorescence (RF), we implement the cross-polarization scheme \cite{kuhlmann_dark-field_2013} to suppress the resonant laser scattering with an extinction ratio of 30 dB. This suppression is not ideal and is crippled by two factors: sample vibrations and the curved sample surface (SIL). In principle, the sample vibrations, on the order of $\sim1\mu$m, in the sample plane should not change the laser scattering if the sample is flat. However, the hemispherical shape of SILs change the polarization of the reflected beam when the illumination depth varies. Since the vibration is random, the laser background thus fluctuate randomly, as shown in ZPL PLE spectra in Fig.~\ref{fig:rple}(c). This explains the deviation of the ZPL PLE linewidth from the 2-level prediction at 6 P$_0$ in Fig. 2(b) of the main text.  
	 
	We find that a proper defocus of the lens used to focus PL into a single-mode fiber helps alleviate this effect and improve the extinction ratio. But strong defocus diminishes the collection efficiency and a trade-off has to be made. We optimize the location of this lens by monitoring the collection efficiency and extinction ratio simultaneously. In the end, we achieved an extinction ratio of 30 dB by sacrificing 20\% collection efficiency, which is 100 times better than the performance at the optimum collection efficiency.

	\section{SECTION 3: PLE, RPLE, lifetime, and PLE linewidth of other GeVs}

	We utilize PLE spectrum at low excitation power to determine the emission linewidth of the GeV color center. Thus, power-dependent PLE is performed on two emitters located in two SILs, as shown in Fig.~\ref{fig:rple}(a) and (b), where the one in SIL B09 is the emitter investigated in the main text. These PLE spectra are collected by monitoring the PSB emissions from the GeVs. The weak side peaks in Fig.~\ref{fig:rple}(b) is caused by the GeVs nearby. Nevertheless, we fit each spectrum with a Lorentzian line and extract the linewidth and central count rate over the excitation power, as shown in Fig. 2(a) of the main text for SIL B09 and Fig.~\ref{fig:rple}(e) for SIL B02. The power dependence follows closely to the model of a coherently drive 2-level system, as expected.  
	
	For the GeV in SIL B09, we also conducted PLE measurements by collecting its ZPL emission, as shown in Fig.~\ref{fig:rple}(c). An evident background rising can be observed as the excitation power increases, which mainly consists of resonant laser leakage due to the finite suppression ratio of $\sim$30 dB. We characterized its excited state lifetime $T_1$ by performing time-resolved measurements, as shown in Fig.~\ref{fig:rple}(d). Here, the resonant excitation beam is modulated by an electro-optic modulator (EOM) with an ON-OFF ratio of 23 dB and a switching time of 300 ps. We collect the PSB emission and focus on the the falling edge of the excitation pulse. The lifetime $T_1$ extracted from the three measurements is close and gives an average value of $5.5\pm0.1$ ns that is employed throughout the work.

	By fitting the power-dependent PLE spectra in Fig.~\ref{fig:rple}(b) with a Lorentzian line, we extract the linewidth and resonance count shown in Fig.~\ref{fig:rple}(e). Similar to the power-dependent behavior of the GeV of the main text, the RF counts and linewidth also follow the predictions of a 2-level system, where we obtain a natural linewidth of 50 MHz by excluding the power broadening. This corresponds to a coherence time of $T_2=7.5$ ns, giving $T_2/2T_1\approx0.82$, i.e., Fourier-transform-limited linewidth emission. These values are comparable to those obtained on SiV color centers in diamond, which is about 0.79 \cite{rogers_multiple_2014}.

%	\noindent  \textbf{1. Lifetime T$_1$} \\
%	\noindent\hspace*{0.35cm} Lifetime of the GeV center is determined by time-resolved PL measurement, as shown in Fig.~\ref{fig:combo}(a). A train of 512-nm 80-ps pulses at a repetition rate of 80 MHz (from PiL051X) is focused onto the GeV center. Zero-phonon line (ZPL) emission from the GeV centr is collected by using a combination of a 600$\pm$7 nm band-pass and a 600 nm long-pass edge filter. The data is fitted with a single exponential decay convolving with the measured instrument response function (IRF, grey curve) of the setup, from which we determine the lifetime of the GeV center to be T$_1 = 3.57 \pm 0.03$ ns.\vspace{0.1cm}

	\begin{figure*}[b]
		\includegraphics{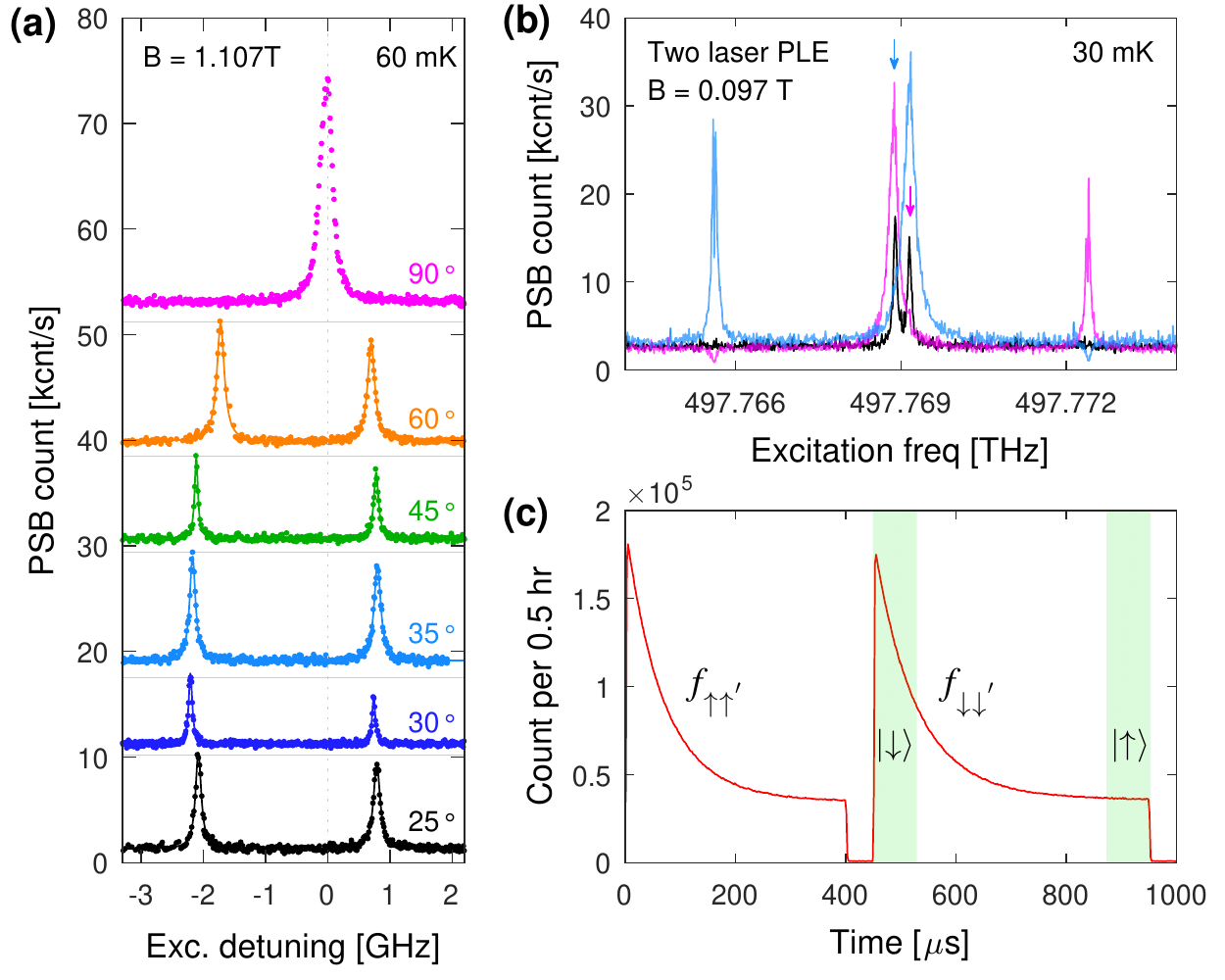}
		\caption{\label{fig:angle} (a) PLE spectra as varying the orientation of the magnetic field with the strength fixed at 1.107 T. The angle is measured with respect to the sample plane. The horizontal grey lines are zero reference for the vertically shifted PLE spectrum. Zero detuning: 497.7698 THz. (b) Two-laser PLE. The black curve is the single-laser PLE taken at 10 nW resonant power and 50 uW green power for reference. The 50 uW green power can induce a weak PL from the GeV. The blue and magenta curves are the two-laser PLE with a laser (2 nW) lock at either low- or high-energy transition (indicated by the arrows) while scan a second laser of 50 nW. The green power is 1 $\mu$W. We note that the dips when both lock and sweep lasers address the same cross transition is caused by spin pumping effect. (c) Time-resolved PL during spin pumping and readout by addressing the spin-selective transitions $f_{\uparrow\uparrow^\prime}$ and $f_{\downarrow\downarrow^\prime}$, respectively. The magnetic field is held at 1.107 T along $0^\circ$. The shaded regions indicate the readout windows of length $\Delta t$ for state $\left|\downarrow\right\rangle$ and $\left|\uparrow\right\rangle$, respectively. 
		}
	\end{figure*}
	
	\section{SECTION 4: Field-direction dependent PLE}

	When placing the GeV in an external magnetic field, its spin degree of freedom can be revealed, as shown in Fig. 1(a) of the main text. Due to the different spin-orbit coupling and different susceptibilities to the strain environment of the ground and excited states, their $g$-factors are not the same, allowing for spin-selective transitions $f_{\uparrow\uparrow^\prime}$ and $f_{\downarrow\downarrow^\prime}$ with $\left| f_{\uparrow\uparrow^\prime} - f_{\downarrow\downarrow^\prime} \right| \propto \delta g$. Since these $g$ tensors are anisotropic, $\delta g$ acquires anisotropy too and gives rise to the field-orientation dependent splitting between $f_{\uparrow\uparrow^\prime}$ and $f_{\downarrow\downarrow^\prime}$, as shown in Fig.~\ref{fig:angle}(a). Interestingly, when the field is perpendicular to the sample plane, the splitting vanishes, implying $\delta g\approx0$ along this direction. 
	
	In all single-laser PLE spectra, as varying the field strength and orientations, we are not able to observe the cross transitions $f_{\uparrow\downarrow^\prime}$ or $f_{\downarrow\uparrow^\prime}$ but only the cycling transitions $f_{\uparrow\uparrow^\prime}$ and $f_{\downarrow\downarrow^\prime}$. Although stronger green laser is helpful to shuffle the population in the spin ground states, it cannot make the cross transitions visible in PLE spectra, as shown by the black curve in Fig.~\ref{fig:angle}(a). The disappearance of the cross transitions $f_{\uparrow\downarrow^\prime}$ or $f_{\downarrow\uparrow^\prime}$ indicates that the orientation of the spin quantization axis of the ground and excited states must be very close. Together with the orthogonal orbital part of the wave function, as granted by the low-strain environment, the probability of cross transitions is very low. But once it is pumped, the population will immediately relax to the other spin ground state with a descent spin pumping efficiency. 
	
	To find the frequencies of cross transitions, we conduct two-laser PLE scans, as shown in Fig.~\ref{fig:angle}(b). Here we lock one laser at a cycling transition while scanning a second laser to acquire the spectrum. The sweep laser is 25 times stronger than the lock laser, partially compensating the low transition probability of cross transitions. This lead to an evident coherent population trapping (CPT) at either cross transition and implies that in spite of the low-strain environment that may cripple the microwave coupling at transition $f_{\downarrow\uparrow}$, two-photon optical process can still be employed to manipulate the ground spin state.

	\begin{figure*}[b]
		\includegraphics[scale=0.9]{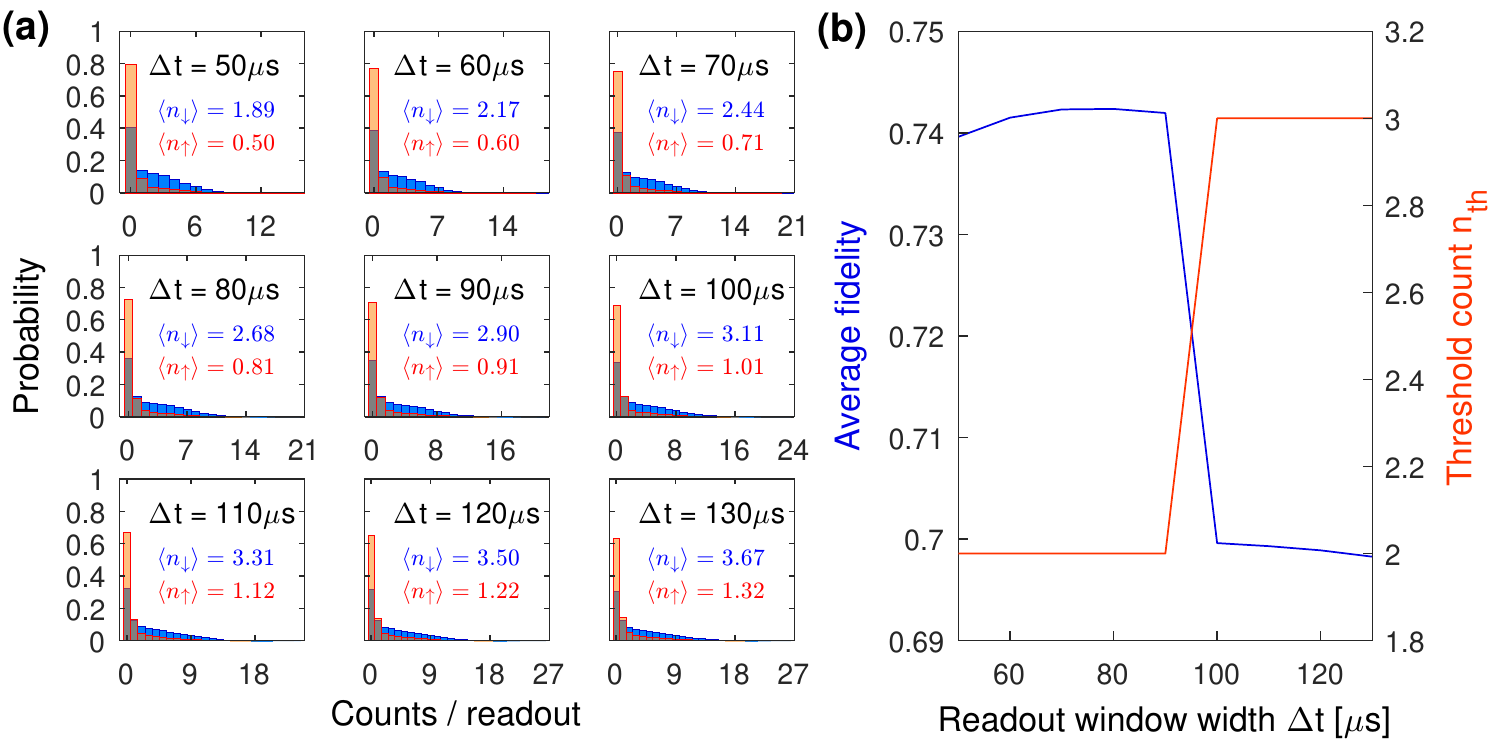}
		\caption{\label{fig:ssr} (a) Readout photon statistics as varying the readout window width $\Delta t$. Red bar corresponds to reading $\left|\uparrow\right\rangle$ state while blue for $\left|\downarrow\right\rangle$ state. $\langle n_\downarrow\rangle$ and $\langle n_\uparrow\rangle$ are the average photon numbers acquired during the readout. (b) Readout fidelity $\mathcal{F}$ and threshold count $n_\text{th}$ extracted from the readout statistics in (a). See the main text for the evaluation formula.  
		}
	\end{figure*}

	\section{SECTION 5: Single-shot readout – threshold $n_\text{th}$ and readout fidelity $mathcal{F}$}

	The single-shot readout of GeV spin state demonstrated in Fig.~2(d) of the main text is performed in 1.107 T magnetic field along the sample plane. At the beginning, the spin is initialized to $\left|\downarrow\right\rangle$ state by pumping $f_{\uparrow\uparrow^\prime}$ transition for 400 $\mu$s. Then $\left|\uparrow\right\rangle$ state is read out by addressing $f_{\downarrow\downarrow^\prime}$ transition for 500 $\mu$s, as shown in Fig.~\ref{fig:angle}(c). To evaluate the readout fidelity, two readout windows of the same length $\Delta t$ are employed for reading $\left|\downarrow\right\rangle$ and $\left|\uparrow\right\rangle$ state, respectively, as shown in Fig.~\ref{fig:angle}(c). We statistics the photon numbers $n_\downarrow$ ($n_\uparrow$) acquired in the readout window by repeating the process for $1.8\times10^6$ times. After normalizing to the total number of occurrences, we obtain the probability distribution $p_\downarrow$($n_\downarrow$) and $p_\uparrow$($n_\uparrow$) to read $n_\downarrow$ and $n_\uparrow$ photons during the $\Delta t$ reading period, as shown in Fig.~\ref{fig:ssr}(a). The photon number of equal probability, i.e., $p_\downarrow$($n_\text{th}$) = $p_\uparrow$($n_\text{th}$), sets the threshold $n_\text{th}$ to determine the final spin state:   $\left|\downarrow\right\rangle$ if $n \geq n_\text{th}$ and $\left|\uparrow\right\rangle$ if $n < n_\text{th}$. The readout fidelities $\mathcal{F}_\downarrow$ and $\mathcal{F}_\uparrow$ are evaluated via $\sum p_\downarrow(n_\downarrow \geq n_\text{th})$ and $\sum p_\uparrow(n_\uparrow < n_\text{th})$, respectively. By varying the readout window length $\Delta t$, we maximize the average fidelity $\mathcal{F} = (\mathcal{F}_\downarrow + \mathcal{F}_\uparrow)/2$ by choosing a time window of 80 $\mu$s, as shown in Fig.~\ref{fig:ssr}(b). This leads to the optimum readout fidelity of $\mathcal{F} =0.74$ and the photon statistics shown by Fig. 2(d) of the main text. Currently, the performance is limited by the poor spin pumping efficiency as shown in Fig.~\ref{fig:angle}(c), where only 83\% of population is pumped into either spin states via resonant driving. This can be improved by aligning the magnetic field to the c3 axis of the GeV to increase the spin relaxation time. 
	Alternatively, cavity-QED technique can be explored to enhance readout fidelity by sparing the requirement on magnetic field direction alignment \cite{sun_single-shot_2016}.

	\section{SECTION 6: Rabi oscillations – model}	
	
	Since orbitals UB and UB$^\prime$ are energetically far from orbitals LB and LB$^\prime$ (see Fig.~1(a) of the main text), the coherently driven transition LB$\leftrightarrow$LB$^\prime$ can be legitimately modeled as a driven 2-level system, where the time evolution of the density matrix follows the master equation \cite{loudon_quantum_2000}
	\begin{align}
	\label{eqn:2lv}
	&\frac{d}{dt} \left(\begin{array}{c}
	\tilde{\rho}_\text{11} \\ \tilde{\rho}_\text{22} \\ \tilde{\rho}_\text{12} \\ \tilde{\rho}_\text{21}
	\end{array}\right) =
	\left(
	\begin{array}{cccc}
	0 & 2\gamma_\text{sp} & i\Omega/2 & -i\Omega/2 \\
	0 & -2\gamma_\text{sp} & -i\Omega/2 & i\Omega/2 \\
	i\Omega/2 & -i\Omega/2 & -1/T_2 & 0 \\
	-i\Omega/2 & i\Omega/2 & 0 & -1/T_2
	\end{array}\right) \left( 
	\begin{array}{c}
	\tilde{\rho}_\text{11}\\
	\tilde{\rho}_\text{22}\\
	\tilde{\rho}_\text{12}\\
	\tilde{\rho}_\text{21}
	\end{array}\right)
	\end{align}
	where $\tilde{\rho}_{ij}$ is the density matrix element of the 2-level in the rotation frame, $\Omega$ is the Rabi frequency, $\omega_0$ is the transition frequency of the two-level system, $\omega$ is the driving frequency,  $\gamma_\text{sp}$ is the spontaneous decay rate of probability amplitude (fixed to half of lifetime $T_1$), and $\gamma$ is the total dephasing rate that includes both $\gamma_\text{sp}$ and pure dephasing $\gamma^\prime$. Since the driving field is periodically on and off, we assign two Rabi frequencies $\Omega_\text{ON}$ and $\Omega_\text{OFF}$ to the above equation to describe the time evolution of the population during ON and OFF period, respectively. In experiments, we used a Mach-Zehnder fiber-based EOM to modulate the resonant excitation beam, with a limited ON-OFF ratio of 23 dB. There is always some excitation power transmitted through the EOM even when it is off, i.e., $\Omega_\text{OFF}\neq0$.
	
\begin{figure*}[b]
		\includegraphics{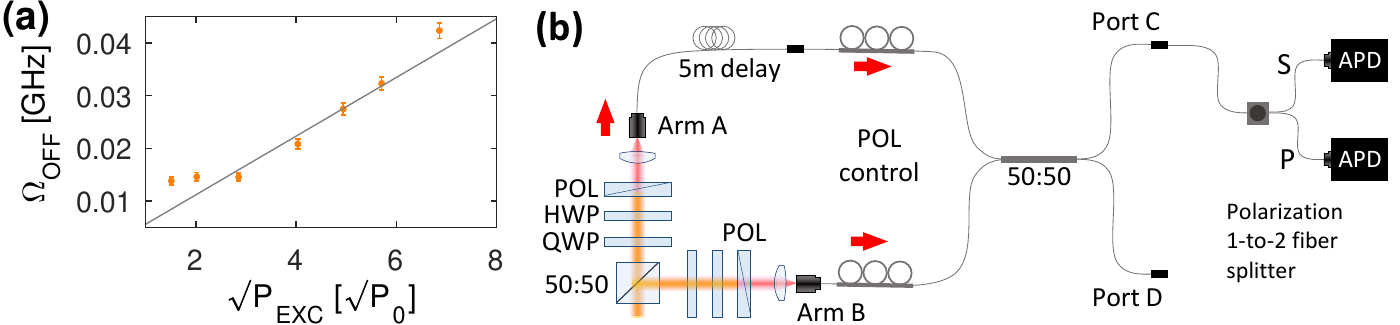}
		\caption{\label{fig:rboff} (a) Rabi frequency $\Omega_\text{OFF}$ when the EOM is off during Rabi oscillation measurements. The finite extinction ratio lets some of the driving field pass through the EOM, giving rise to a low value $\Omega_\text{OFF}$. The two data points at the low end of excitation power is too weak to be extracted accurately.Thus we exclude these two data points for the linear fit. (b) Schematic for aligning the polarizations of interfering photons for $g_\parallel^{(2)}(\tau)$ and $g_\perp^{(2)}(\tau)$ in Fig.~4 of the main text. An extra polarization-dependent 1-to-2 fiber splitter is added to the end of the TPI setup. By blocking either detection arm and minimizing the count rate at the same or different APDs, we can align the polarizations to be parallel or perpendicular to each other. 
		}
	\end{figure*}

	By numerically solving the above equations and fit the data shown in Fig.~3(a) of the main text, we extract $\Omega_\text{ON}$, $\Omega_\text{OFF}$ and $T_2$ over different excitation powers. Parameter $\Omega_\text{ON}$ and $T_2$ have been shown and analyzed in Fig.~3(c) and (d) of the main text. $\Omega_\text{OFF}$ is plotted in Fig.~\ref{fig:rboff}(a) below. It increases linearly over the square root of the excitation power, as expected.

	\section{SECTION 7: Polarization alignment for TPI experiments}	
	
	When conducting the two-photon interference (TPI), we adjust the polarizations of interfering photons to realize distinguishable or indistinguishable configurations, as shown in Fig. 4 of the main text. This is done by tuning the polarization controls attached to the single-mode fibers. Before any adjustment, we connect one output of the 50:50 non-polarizing fiber splitter (e.g., port c) to the 1-to-2 polarizing fiber splitter (from OZoptics), as shown in Fig.~\ref{fig:rboff}(b). Similar to the bulk PSB cube, this fiber splitter sorts the incoming photons into two outputs based on its polarizations. We monitor the intensities of outputs with avalanche photodiodes (APDs).
	
	Block Arm B, open Arm A, and adjust the polarization control on Arm A to minimize the intensity at S-APD. Now the polarization of Arm A is set to deliver P-polarized light at the 1-to-2 fiber BS. Similarly, open Arm B, block Arm A, and minimize the S-polarized counts again by tuning Arm B polarization. This ensures the aligned polarization of Arms A and B. We note that, during the alignment, it is critical to keep all fibers static (fixed), since any strain on the SM fiber can change the polarization of the light at the output. To obtain cross polarization between Arm A and B, we minimize the count rate at S-APD and P-APD by tuning the polarization of Arm A and B, respectively.

	%\begin{figure*}[b]
	%	\includegraphics{./Figure/FigS8_v1_0_TPI_fitlers.pdf}
	%	\caption{\label{fig:filter} Time-resolved accumulated photon counts during the pulsed TPI measurements for cross (a) and parallel (b) polarization, respectively. The red, blue, and black curves are the counts accumulated on CH0, CH1, and the sum of the two, respectively. The shaded regions mark the six time filters, three for each detection channel (color matched). The photon events within these filters are extracted and used to calculate the TPI correlations as shown Fig.~4(c) and (d) of the main text. }
	%\end{*}
	
	\section{SECTION 8: TPI - continuous-wave excitation}	
	
	Generally, the TPI photon statistics measured for CW-driven GeV, as shown in Fig. 4(b) of the main text, consists of two parts: the TPI statistics and the correlations involving the resonant laser photons 	
	\begin{equation}
	\label{eqn:2levelpop}
	g^{(2)}(\tau) = \frac{\langle I_1(t)I_2(t+\tau) \rangle}{\langle I_1(t)\rangle \langle I_2(t+\tau)\rangle} =  \frac{\langle (S_1(t) + B_1(t))(S_2(t+\tau) + B_2(t+\tau)) \rangle}{\langle S_1(t)+B_1(t)\rangle \langle S_2(t+\tau) + B_2(t+\tau)\rangle} 
	\end{equation}
	where $S(t)$ and $B(t)$ represent the intensities of signal and laser background, respectively. The subscript 1 and 2 denotes the two detection channels. Assuming no correlations between the signal and the background, i.e., $\langle S\cdot B\rangle = \langle S\rangle\cdot\langle B\rangle$, we obtain
	\begin{equation}
	\label{eqn:g2expand}
	g^{(2)}(\tau) = \frac{\langle S_1(t)S_2(t+\tau)\rangle + \langle B_1(t)\rangle\langle S_2(t+\tau)\rangle + \langle S_1(t)\rangle\langle B_2(t+\tau)\rangle + \langle B_1(t)B_2(t+\tau)\rangle} {\langle S_1(t)\rangle\langle S_2(t+\tau)\rangle + \langle B_1(t)\rangle\langle S_2(t+\tau)\rangle + \langle S_1(t)\rangle\langle B_2(t+\tau)\rangle + \langle B_1(t)\rangle\langle B_2(t+\tau)\rangle}
	\end{equation}
	%g^{(2)}(\tau) = \frac{\langle S_1(t)S_2(t+\tau)\rangle + \langle B_1(t)\rangle\langle S_2(t+\tau)\rangle + \langle S_1(t)\rangle\langle B_2(t+\tau)\rangle + \langle B_1(t)B_2(t+\tau)\rangle} {\{\langle S_1(t)\rangle + \langle B_1(t)\rangle\} \{\langle S_2(t+\tau)\rangle + \langle B_2(t+\tau)\rangle\}}
	
	By switching off the green laser, we quench the RF from the GeV, allowing for the experimental determination of the ratio of $\langle S(t)\rangle$ and $\langle B(t)\rangle$ as $\eta = \langle B \rangle/\langle S\rangle = 1/4$. 
	Plugging it in Eqn.~\ref{eqn:g2expand} gives 
	
%	\begin{equation}
%	\label{eqn:eta}
%	\eta = \frac{\langle B \rangle}{\langle S\rangle} =  \frac{1}{4} 
%	\end{equation}

	\begin{equation}
	\label{eqn:g2perp}
	g^{(2)}_\perp(\tau) = 1 - \frac{1}{(1+\eta)^2} + \frac{T_2 R_2 (T_1^2+R_1^2) g^{(2)}(\tau) + R_1 T_1 [T_2^2 g^{(2)}(\tau + \Delta t) + R_2^2 g^{(2)}(\tau-\Delta t)]}{A(1+\eta)^2} 
	\end{equation}
	where $A=R_2 T_2 (T_1^2+R_1^2 )+R_1 T_1 (T_2^2+R_2^2 )$ is the normalization constant, and $R_1$, $T_1$, $R_2$, and $T_2$ are the intensity reflection and transmission coefficients of BS1 and BS2, respectively. We experimentally verified these values which are are quite close to 0.5. So they are fixed to 0.5 for the fitting. 
	
	Similarly, for the indistinguishable photons, we have
	\begin{equation}
	\label{eqn:g2para}
	g^{(2)}_\parallel(\tau) = 1 - \frac{1}{(1+\eta)^2} + \frac{T_2 R_2 (T_1^2+R_1^2) g^{(2)}(\tau) + R_1 T_1 [T_2^2 g^{(2)}(\tau \Delta t) + R_2^2 g^{(2)}(\tau-\Delta t)] - R_1 T_1 (R_2^2 + T_2^2)V_0 \left|g^{(1)}(\tau)\right|^2}{A(1+\eta)^2} 
	\end{equation}
	where parameter $V_0$ includes all experimental imperfections that destroy the overlap in space or polarization of the two beams at the second BS. By solving the master equation numerically, we can evaluate $g^{(2)}(\tau)$ and $g^{(1)}(\tau)$, respectively, and use the final $g^{(2)}_\parallel(\tau)$ and $g^{(2)}_\perp(\tau)$ to implement the fitting.

	\section{SECTION 9: TPI - pulsed excitation}
	
	To further reduce the statistical contaminations caused by the leaked resonant laser light, we time filtered the detected photon events and post-selected those registered within $\sim3 T_1$ period after the falling edge of the excitation pulse to calculate the correlations.%, as shown in Fig.~\ref{fig:filter}. 
	The extracted correlations based on these filtered photon events are shown in Fig.~4(c) and (d) of the main text.
	
	We also develop a semi-classical model to analyze the pulsed excitation results in Fig.~4(c) and (d) of the main text by following Ref.~\cite{kiraz_quantum-dot_2004}. Considering two photons emitted from a single emitter excited by two pulses with $\delta t$ separation, collecting by two detection arms with a relative delay of $\delta t$, the photon state at the entrance of the second BS (i.e., the 50:50 fiber BS in Fig.~1(a) of the main text) becomes
	\begin{align}
	\label{eqn:psi0}
	\left|\psi_\text{in}(t)\right\rangle =& \frac{1}{2} \big{(} T_1 \left|\phi_1^A,t+\delta t\right> \left|\phi_2^A, t+2\delta t\right> - R_1 \left|\phi_1^B,t\right> \left|\phi_2^B, t+\delta t\right> \notag\\
	& + i\sqrt{T_1 R_1}\left|\phi_1^A, t+\delta t\right>\left|\phi_2^B, t+\delta t\right> + i\sqrt{T_1 R_1}\left|\phi_2^A, t+2\delta t\right>\left|\phi_1^B, t\right>\big{)}
	\end{align}
	where $T_1$ and $R_1$ are the transmission and reflection intensity coefficient of the first BS, $\phi_1$ and $\phi_2$ label the first and second photon emitted within the two excitation pulses, and superscript A and B label the collection arm. Here, the third term gives rise to the HOM interference at $\tau=0$, while the rest three are responsible for the correlation events observed at $\left|\tau\right|=\delta t$ and $2\delta t$. 
	
	\begin{figure}[thbp]
		\centering
		\includegraphics[scale=2]{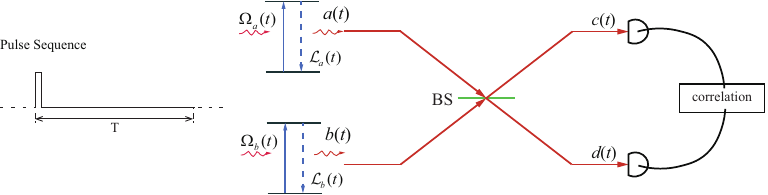}
		\caption{\label{hom} Schematic of HOM interference between two emitters under pulsed resonant excitation with a period of T. The HOM interference is equivalent to the interference of TPI experiment based on a single quantum emitter as explored by the experiment. The BS here corresponds to the 50:50 fiber BS in Fig.~1(a) of the main text. Operator $a(t)$, $b(t)$, $c(t)$ and $d(t)$ are field annihilation operators of different paths, and $\Omega_k(t)$ and $\mathcal{L}_k(t)$ are the Rabi frequency and Lindblad term for emitter $k$.  % The master equations are described in Sec. 6.
		}
	\end{figure}
	
	The correlation from the third term of Eqn.~\ref{eqn:psi0} can be evaluated from the TPI between two independent emitters, as depicted in Fig.~\ref{hom}. Assuming that the BS possesses a transmission and reflection coefficient $T_2$ and $R_2$, we evaluate the second-order optical coherence via
	\begin{align}
	\left<c^\dagger(t) d^\dagger(t+\tau) d(t+\tau) c(t)\right> =&~T_2^2 \left<b^\dagger(t) a^\dagger(t+\tau) a(t+\tau) b(t)\right> + R_2^2 \left<a^\dagger(t) b^\dagger(t+\tau) b(t+\tau) a(t)\right> \notag\\
	&- R_2 T_2 {\big[} \left<a^\dagger(t) b^\dagger(t+\tau) a(t+\tau) b(t)\right> + \left<b^\dagger(t) a^\dagger(t+\tau) b(t+\tau) a(t)\right> {\big]}
	\end{align}
	
	In the far-field regime, the field annihilation operator $a$ and creation operator $a^\dagger$ can be replaced by the dipole operators $\sigma_{ge}$ and $\sigma_{eg}$ via source-field relationship
	\begin{align}
	\label{eqn:source}
	a(t) = A(\vec{r}) \cdot \sigma_\text{ge}(t-\left|\vec{r}\right|/c)
	\end{align}
	where $A(\vec{r})$ is a time-independent factor describing the spatial distribution of the field. Since the two emitters ($a$ and $b$) share the same expectation value and coherence function, we set $\left<b(t)\right> \approx \left<a(t)\right>$ \cite{kiraz_quantum-dot_2004}. Thus, 
	%By substituting the photon annihilation ($a$) and creation ($a^\dagger$) operator with the corresponding dipole operators $\sigma_{ge}$ and $\sigma_{eg}$ via Eqn.~\ref{eqn:source}, we obtain
	\begin{align}
	\label{eqn:3rd}
	\left<c^\dagger(t)d^\dagger(t+\tau)d(t+\tau)c(t)\right> = \left|A(\vec{r})\right|^4 \cdot [(R_2^2 + T_2^2) \left<\sigma_{ee}(t)\right> \left<\sigma_{ee}(t+\tau)\right> - 2R_2 T_2 \left|\left<\sigma_{eg}(t+\tau)\sigma_{ge}(t)\right>\right|^2]
	\end{align}
	%Neglecting the normalization constant $\left|A(\vec{r})\right|^4$, 
	
	Now we define
	\begin{subequations}
		\begin{align}
		G_{\perp}^{(2)}(\tau) &= \lim_{T\to\infty} \int_{0}^{T} dt (R_2^2+T_2^2) \left<\sigma_{ee}(t)\right> \left<\sigma_{ee}(t+\tau)\right>,\\
		G^{(1)}(t,\tau) &= \sqrt{2R_2T_2}\left<\sigma_{eg}(t+\tau)\sigma_{ge}(t)\right>,\\
		G_{\parallel}^{(2)}(\tau) &= G_{\perp}^{(2)}(\tau) - \lim_{T\to\infty} \int_{0}^{T}dt \left|G^{(1)}(t,\tau)\right|^2
		\end{align}
	\end{subequations}
	where the integration over $t$ covers the entire experiment acquisition time. In simulations, we find that a few periods integration is good enough and we set four periods for evaluations. The final TPI visibility can be calculated
	\begin{align}
	\label{eqn:vis}
	\mathcal{I} = \dfrac{\int d\tau G_{\perp}^{(2)}(\tau)-\int d\tau G_{\parallel}^{(2)}(\tau)}{\int d\tau G_{\perp}^{(2)}(\tau)}
	\end{align}
	where the integration over $\tau$ covers the central correlation peak around $\tau=0$. 
	
	Following the similar derivation for Eqn.~\ref{eqn:3rd}, the correlation of the last term in Eqn.~\ref{eqn:psi0} ($\left|\phi_2, t+2\delta t\right> \left|\phi_1,t\right>$) can be derived as 
	\begin{align}
	G_{2\delta t, \parallel}^{(2)}(\tau) = G_{2\delta t, \perp}^{(2)}(\tau) = [G_{\perp}^{(2)}(\tau+2\delta t) + G_{\perp}^{(2)}(\tau-2\delta t)]/2
	\end{align}

	Regarding the rest two terms in Eqn.~\ref{eqn:psi0}, $\left|\phi_1,t+\delta t\right> \left|\phi_2,t+2\delta t\right>$ and $\left|\phi_1,t\right> \left|\phi_2, t+\delta t\right>$, both contribute to correlations at $|\tau|=\delta t$. Since the correlated photons come from the same arm, the correlation events are essentially HBT type. 
	Since $\delta t \gg T_1$, no auto field-field interference takes place. The first-order coherence must vanish, i.e.,  $\int d\tau\int dt \left|G^{(1)}(t,\tau)\right|^2 = 0$, leading to
	\begin{align}
	\left<c^{\dagger}(t)d^{\dagger}(t+\tau)d(t+\tau)c(t)\right> = \left|A(\vec{r})\right|^4 \cdot R_2 T_2 \left<\sigma_{eg}(t)\sigma_{ee}(t+\tau)\sigma_{ge}(t)\right>
	\end{align}
	The final accumulated events are 
	\begin{align}
	G_{\mathrm{HBT}}^{(2)}(\tau\pm\delta t) = \lim_{T\to\infty}\int_0^T dt R_2T_2 \left<\sigma_{eg}(t\pm\delta t) \sigma_{ee}(t+\tau\pm\delta t) \sigma_{ge}(t\pm\delta t)\right>
	\end{align}
	Summing up all contributions, we obtain the final cross and parallel correlation function $g_{\perp}^{(2)}(\tau)$ and $g_{\parallel}^{(2)}(\tau)$
	\begin{align}
	g_{\perp}^{(2)}(\tau) &= N\left\lbrace 4G_{\mathrm{HBT}}^{(2)}(\tau+\delta t) + 4G_{\mathrm{HBT}}^{(2)}(\tau-\delta t) + 2G_{\perp}^{(2)}(\tau) + G_{\perp}^{(2)}(\tau+2\delta t) + G_{\perp}^{(2)}(\tau-2\delta t) \right\rbrace  \\
	g_{\parallel}^{(2)}(\tau) &= N\left\lbrace 4G_{\mathrm{HBT}}^{(2)}(\tau+\delta t) + 4G_{\mathrm{HBT}}^{(2)}(\tau-\delta t) + 2G_{\parallel}^{(2)}(\tau) + G_{\perp}^{(2)}(\tau+2\delta t) + G_{\perp}^{(2)}(\tau-2\delta t) \right\rbrace  \label{eqn:gpara}
	\end{align}
	where $N$ is a normalization constant.

	\begin{figure}[thbp]
		\centering
		\includegraphics[scale=1]{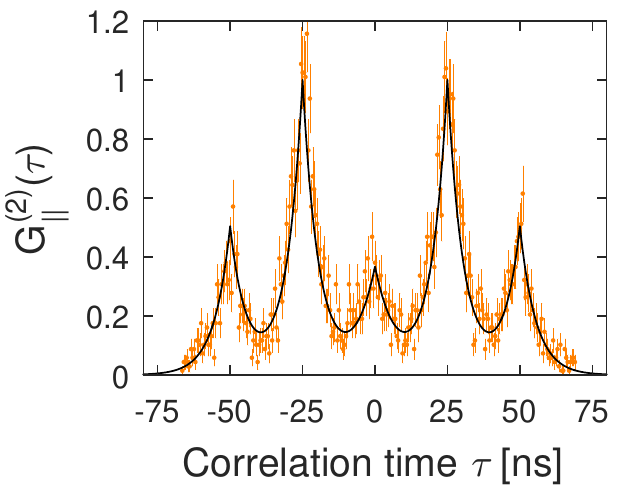}
		\caption{\label{fig:simu} Fitting of the TPI result with the model described by Eqn.~\ref{eqn:gpara} and Eqn.~\ref{eqn:fit}. Orange dots are the data, the same as Fig.~4(d) of the main text. The black curve is the fitting with $T_2=9.68\pm0.65$ ns and $\mathcal{V}=0.649\pm0.036$. The excited state lifetime is fixed to 5.5 ns during the fitting.}
	\end{figure}
	
	%We compared two cases: including and excluding the excitation period in correlation evaluations, as shown in Fig.~\ref{fig:simu}(a) and (b), respectively.	The fact that the inclusion of the excitation period [Fig.~\ref{fig:simu}(a)] results in rounder correlation peaks than those of only considering post-excitation period [Fig.~\ref{fig:simu}(b)] implies that the lifetime length of the excitation $\pi$-pulse is still too long to suppress the multi-photon emissions from the quantum emitter. In data parsing, since we only take the photons emitted after the excitation period into account for correlation extractions [see Fig.~\ref{fig:filter}], the simulation in Fig. \ref{fig:simu}(b) thus better matches to the data as expected. Nevertheless, both simulations exhibit a dip at $\tau=0$, reflecting the indistinguishability of photons. But the dip is not wide enough to diminish the entire central peak due to the finite coherence of the interfering photons ($T_2<2T_1$). The fact that no dip appears in the data implies the imperfection of our TPI experiment. One possible source can be the non-ideal spatial or temporal overlap of photons, which is usually parameterized by a factor $\mathcal{V}$ to the coherence term $|g^{(1)}(\tau)|^2$ 
	The imperfection of experiment, such as non-ideal spatial/temporal overlap of photons or the resonant laser leakage (due to the finite suppression), can be parameterized by a factor $\mathcal{V}$ to restrain the interfering effect
	\begin{align}
	\label{eqn:fit}
	G_{\parallel}^{(2)}(\tau) &= G_{\perp}^{(2)}(\tau) - \mathcal{V}\lim_{T\to\infty}\int_{0}^{T}dt \left|G^{(1)}(t,\tau)\right|^2,
	\end{align}
	where $\mathcal{V}$ takes a value between 0 and 1. 
	By fitting the data with Eqn.~\ref{eqn:fit}, we extracted $T_2=9.68\pm0.65$ ns and $\mathcal{V}=0.649\pm0.036$, as shown in Fig.~\ref{fig:simu}. Here, the value of $T_2$ is much higher than the expectation of 3.5 ns at 39 $P_0$ if the emitter is CW driven at the same power level [see Fig.~3(c) of the main text]. This might also relate to the laser induced decoherence, which is active when the laser is on and is suppressed effectively when the laser is off. Thus, by selecting the photons emitted after switching off the laser, we obtained a significantly longer coherence time.

	% \bibliography{./Bibliography/GeV_TPI_supp_v1}
	%